\documentclass[a4paper,12pt]{article}
\usepackage{amssymb}
\usepackage{psfig}
\usepackage{here}
\newcommand{\beq}{\begin{equation}}
\newcommand{\eeq}{\end{equation}}
\newcommand{\beqa}{\begin{eqnarray}}
\newcommand{\eeqa}{\end{eqnarray}}

\def\un{\underline}

\title{\textbf{Single chain elasticity and thermoelasticity of polyethylene}}
\author{\centerline{John T. Titantah$^{1,2}$\thanks{Present address: Theoretical Study of Matter, 
Department of Physics, University of Antwerpen, Groenenborgerlaan 171, 2020 Antewerpen (Belgium)}~, 
Carlo Pierleoni$^1$ and Jean-Paul Ryckaert $^2$}
\\ $^1$ INFM and Dipartimento di Fisica, Universit\`a degli studi, \\Via Vetoio, I-67100 L'Aquila, Italy 
\\ $^2$ Unit\'e de Physique des Polym\`eres, CP 223, Universit\'e Libre de Bruxelles,
\\Bd du Triomphe, B-1050 Brussels, Belgium}

\begin{document}

\maketitle

\section*{Abstract}
Single-chain elasticity of polyethylene at $\theta$ point up to $90\%$ of
stretching with respect to its contour length is computed by Monte-Carlo
simulation of an atomistic model in continuous space. The elasticity law
together with the free-energy and the internal energy variations with
stretching are found to be very well represented by the wormlike chain model up
to $65\%$ of the chain elongation, provided the persistence length is treated
as a temperature dependent parameter. Beyond this value of elongation simple
ideal chain models are not able to describe the Monte Carlo data in a
thermodynamic consistent way. This study reinforces the use of the wormlike
chain model to interpret experimental data on the elasticity of synthetic
polymers in the finite extensibility regime, provided the chain is not yet in
its fully stretched regime. Specific solvent effects on the elasticity law and
the partition between energetic and entropic contributions to single chain
elasticity are investigated.

\newpage

\section{Introduction}

When a single linear chain of N segments is stretched at both ends by equal and
opposite forces $\pm {\un f}$, the average end-to-end vector $<{\un R}>$ lies
in the direction of the force and its magnitude $R$ is related to $f$ by an
elasticity law or equation of state (EOS) that is usually casted in the
dimensionless form \beq \frac{f\ell}{k_BT}=w(R,N,T)\label{eq:EOS} \eeq where
$k_B$ is the Boltzmann constant, T the absolute temperature, $\ell$ a
characteristic length of the unstressed chain and $w$ a dimensionless function.

At very low force intensities the equation of state  follows the simple linear
Hooke's law \cite{dG79}
\begin{equation}
\frac{f R_0}{k_BT}=3\frac{R}{R_0}
\label{eq:lin}
\end{equation}
where $R_0$ is the end-to-end vector (in the mean square sense) of the polymer
in its unstressed state. For somehow larger forces ($fR_0/k_BT\geq2$) the so
called ``Pincus'' regime is entered\cite{WLK81,P76}
\beq 
\frac{f R_0}{k_BT} \sim \left(\frac{R}{R_0}\right)^\gamma
\label{eq:pincus} 
\eeq
where $\gamma=1,~3/2$ for $\theta$ and good solvent conditions respectively
\cite{dG79}. Note that for $\theta$ chains the two regimes merge in a single
one. The Pincus regime is abandoned in favor of the so called ``Finite
Extensibility'' regime (FE) when the applied force is further increased beyond
$f b_k/k_BT=1$ where $b_k$, the Kuhn segment of the polymer, is a measure of
the local rigidity of the chain. In this FE regime, the EOS depends on the
microscopic details of the specific chain being stretched and, for long enough
chains, the EOS can be generally written
\beq
\frac{f b_k}{k_BT}=w(X,T)
\label{eq:finex}
\eeq
where $X=R/L_c$ and $L_c$ is the chain contour length.

Elasticity laws of real chains in this regime are usually discussed in terms of
simple ideal models like the Freely Jointed chain (FJC), the Wormlike chain
(WLC) and some ad hoc extensions (elastic FJC and elastic WLC)
\cite{Odijk95,SFB92,MS95,Merkel01,JNO00}. The great success of the WLC in
describing the elasticity of double-stranded DNA and of other biopolymers
\cite{MS95} suggests the existence of some form of universality in the FE
regime also. In general the ideal chain models predict an EOS of the form
(\ref{eq:finex}) with a temperature independent right hand side $w=w(X)$
\beq
\frac{f b_k}{k_BT}=w(X)
\label{eq:models}
\eeq
In eq. (\ref{eq:models}), $b_k$ is usually treated as a constant, the force
being thus proportional to the temperature at any fixed relative extension of
the chain, in good agreement with the well known thermo-elasticity properties
of rubbers. However, the variation with temperature of the size of the polymer
coil in a melt \cite{Flory,BRB91} implies a temperature dependence of
$b_k=R_0^2/L_c$, meaning that the local rigidity of the chain is
affected by the temperature. In the context of single chain elasticity, the
ideal chain models EOS eq. (\ref{eq:models}) are used to interpret the
experimental data on the $X$ dependence of the force. The Kuhn segment $b_k$
and the contour length $L_c$ are considered as fitting parameters in the
analysis, but the temperature dependence of $b_k$ cannot be inferred since
experiments are usually performed isothermally at a unique
temperature\cite{SFB92,OH99,CWA99,YTF00}. When attention is payed to energetic
aspects of the stretching, the temperature dependence of $b_k$ becomes a
sensible point. If the chain EOS follows eq. (\ref{eq:models}) with $b_k$
constant, thermoelasticity predicts that the internal energy of the chain does
not change with stretching and equivalently, that the internal force has a pure
entropic character ($f=f_S$). The same EOS eq. (\ref{eq:models}) with a
temperature dependent $b_k$ predicts that the internal energy change due to
isothermal stretching is proportional to the corresponding free energy change
over the whole range of stretching and the ratio $f_S/f (\neq 1)$ is
independent of stretching. Therefore, ideal models with temperature dependent
Kuhn segment have the whole thermodynamic complexity, and are very promising
candidates to interpret real chain behaviors.

The temperature dependence of the Kuhn segment has been much investigated
experimentally for a large class of synthetic polymers \cite{Flory} and more
particularly for polyethylene (PE) (see ref \cite{BRB91} and references
therein). It has been obtained from the temperature dependence of the radius of
gyration of the chain in melts, from intrinsic viscosity measurements of the
polymer in various $\theta$ point solvents or from thermoelasticity
measurements on polymer networks. For PE, the chain dimension in the $\theta$
state is found to increase with decreasing temperature. On the basis of
Rotational Isomeric State (RIS) model calculations, this trend has been related
to the increase of trans population (respect to gauche) of the dihedral angles
of the chain backbone as temperature drops \cite{AF70} which leads to a local
straightening of the chain reflected by an increase of the length of the Kuhn
segment.

On this basis it is tempting to test more deeply the validity of ideal models to
represent real chain thermo-elasticity by measuring single chain elasticity law
for synthetic polymers. Experimentally, the interpretation of such elasticity
data is made difficult by the limited signal to noise ratio and the limited
range of forces accessible which hamper an accurate data fitting
\cite{OH99,JNO00,Merkel01}.

A possible alternative route to investigate single chain elasticity is provided
by numerical simulation of theoretical models, either coarse grained or
atomistic \cite{CB95,KB96,PAR97,TPR99,MM99,KGZO99,HG00}. In the present work we
use the Monte-Carlo (MC) simulation approach on a well defined polymer
atomistic model to establish with accuracy the chain elasticity law over a wide
range of extension $(0<R/L_c<0.9)$ including the finite extensibility regime
and the corresponding change in the chain internal energy with stretching. Our
aim is twofold: i) to assess the validity of an EOS of the type of eq.
(\ref{eq:models}) definitely simpler than the general form of eq.
(\ref{eq:finex}); ii) to test the applicability of simple ideal models to
represent MC data. To follow this program, we consider the synthetic polymer
with the simplest microscopic structure, namely PE and we adopt an atomistic
model in continuous space. To access both weak and high stretching regimes with
good statistical accuracy, we exploit the configurational bias MC methodology
in combination with biased fixed-$f$ ensemble sampling techniques.

The thermoelasticity of PE has been the subject of a previous Monte Carlo study
for a lattice model \cite{CB95} but that work is focused on the study of the
solvent effect more than on the EOS and the test of ideal chain models.

The paper is organized as follows: in the next section, we review the
theoretical framework of single chain thermo-elasticity and its formal
statistical mechanics basis. With the aim of performing single chain
simulations, we review the theoretical status of a single chain models meant to
be representative of a chain embedded in a solvent (or in a bath of similar
chains) for which intramolecular potentials have a free energy character. In
section III, we give for completeness the adopted PE model, namely the
united-atom model devised originally by A. Sariban et al. \cite{SBR92,SMB94}.
We give in section IV some details about the PE simulations performed within
the fixed force ensemble, the results of which can be used to get also averages
in the fixed end-to-end vector ensemble. In section V, we analyze the results
for PE chains, most of them being relative to $\theta$ point (melt state) at
400K while a few data are presented for PE in good solvent. We discuss the free
energy and internal energy evolutions with stretching in the FE regime, looking
at the partition between the various energy contributions. We analyze the
adequacy of the FJC and WLC models to represent the thermo-elastic behavior of
PE. Section VI is concerned with some results and a short discussion on the
effects of excluded volume on the elasticity law of PE. Section VII gathers our
conclusions.

\section{Theoretical framework}

\subsection{Thermodynamics}

The central quantity in single chain thermo-elasticity is the
chain Helmholtz free energy $A(R,T)$ the total differential being
\begin{equation}
 dA=-SdT+{\un f}\cdot d{\un R}.
\end{equation}
The corresponding single chain internal energy is then given by
\begin{equation}
 E({\un R},T)=\left({\partial \beta A({\un R},T)\over \partial
 \beta}\right)_{\un R}
 \label{eq:ER}
\end{equation}
where $\beta=1/k_BT$. The internal force existing within a stretched
chain with end-to-end vector $\un R$ can naturally be split into an
entropic (${\un f}_S$) and an energetic part (${\un f}_E$)according to
\begin{equation}
 {\un f}({\un R},T)=\left({\partial A({\un R},T)\over \partial
 {\un R}}\right)_T={\un f}_E+{\un f}_S
\label{eq:fder}
\end{equation}
 where
\begin{equation}
 {\un f}_E=\left({\partial E({\un R},T)\over \partial
 {\un R}}\right)_T=-T^2\left({\partial \left[{\un f}({\un R},T)/ T\right]\over \partial
 T}\right)_{\un R}
\label{eq:fe}
\end{equation}
and
  \begin{equation}
 {\un f}_S=-T\left({\partial S({\un R},T)\over \partial
 {\un R}}\right)_T=T\left({\partial {\un f}({\un R},T)\over \partial
 T}\right)_{\un R}
\label{eq:fs}
\end{equation}

\subsection{Statistical Mechanics}

The system under study is a single homo-polymer chain, with degree of
polymerization $N$, which is surrounded by a bath of similar (melt case) or
dissimilar (solvent) molecules (note that in the following $N$ will always
represent the number of skeletal bonds in the linear chain). To simplify
notations, the $N$ dependence of single chain quantities will not be made
explicit in the following formal developments.

The link between the Helmholtz free energy $A({\un R},T)$ and
statistical mechanics is
\begin{equation}
 A({\un R},T)=-k_BT\log Z_R({\un R},T)
 \label{eq:FE}
\end{equation}
where the partition function $Z_R({\un R},T)$, appropriate to a
fixed-$R$ ensemble (hence the $R$ index to emphasize the nature of
constraint), is expressed as
\begin{equation}
Z_R({\un R},T)=\int d\!{\un\Gamma}_1 \delta\left({\un r}_N -{\un
r}_0-{\un R}\right)\exp\left(-\beta \tilde U({\un\Gamma}_1
;T,\rho)\right) \label{eq:ZR}
\end{equation}
where ${\un \Gamma}_1$ is the set of cartesian coordinates $\left(\un
r_i\right)_{i=0,N}$ of the $N+1$ point particles of the chain. In this
equation, $\tilde U({\un \Gamma}_1)$ is the effective potential (or potential
of mean force) characterizing the chain embedded in a bath consisting of the
other chains of the melt or the solvent in an infinite dilution situation. The
bath is characterized by a temperature $T$ and melt or solvent density $\rho$
and therefore the effective potential $ \tilde U({\un\Gamma}_1)$ originates
from an integration over the (infinitely large number of) bath degrees of
freedom $\!{\un\Gamma}_2$, namely
\begin{equation}
\exp\left(-\beta \tilde U({\un\Gamma}_1;T,\rho)\right) \equiv \int
d{\un\Gamma}_2 \exp\left(-\beta U({\un\Gamma}_1,{\un\Gamma}_2)\right)
\label{eq:meanpot}
\end{equation}
where $U({\un\Gamma}_1,{\un\Gamma}_2)$ is the bare total potential of
the chain + bath system.

We further note the well known equivalence
\begin{equation}
Z_R({\un R},T)=Z(T) W^{0}(R,T) \label{eq:histo}
\end{equation}
where $Z(T)=\int d{\un R} Z_R({\un R},T)$ is the partition function of
the free chain in the bath at temperature T and $W^{0}({\un R},T) d{\un
R}$ is the probability to find the end-to-end vector of that free chain
in the range between ${\un R}$ and ${\un R} + d{\un R}$.

Any single chain property $O({\un R},T)$ (such as the tensile
internal force or any structural quantity like a bond-alignment
factor) is an average computed over the subset of configurations
of the free chain which satisfy a fixed-$R$ value. More
explicitly, the average of the microscopic quantity $\hat{O}$,
denoted as $<\hat{O}>_R$, reads
\begin{equation}
O({\un R},T) \equiv <\hat{O}>_R = Z_R({\un R},T)^{-1} \int
d{\un\Gamma}_1 \delta\left({\un r}_N -{\un r}_0-{\un R}\right)
\hat{O}({\un\Gamma}_1) \exp\left(-\beta \tilde U({\un\Gamma}_1
;T,\rho_s)\right) \label{eq:averag}
\end{equation}
The corresponding average $O(T)$ for the free chain is related to the fixed-$R$
averages through the relationship
\begin{equation}
O(T)= \int d{\un R}~ W^{0}(R,T)~ O(R,T)
\end{equation}
We emphasize here the special case of the internal energy $E(R,T)$
which can be expressed as an average quantity of the type of
eq.(\ref{eq:averag}). Indeed, due to the free energy nature (temperature
dependency) of the potential, one gets by combining basic eqs.
(\ref{eq:ER}),(\ref{eq:FE}),(\ref{eq:ZR}) and (\ref{eq:meanpot})
the expression
\begin{equation}
E({\un R},T)=<{\partial (\beta \tilde U)\over\partial \beta}>_R
\label{eq:energy}
\end{equation}

\section{Model of polyethylene chain}

Our purpose is to extract the elasticity law of a PE model which is
sufficiently realistic to be pertinent to that specific polymer but which
avoids explicit consideration of the bath degrees of freedom. This is done by
modeling the system of interest at the level of the $\tilde U({\un\Gamma}_1)$
function for which we adopt the Sariban model \cite{SBR92,SMB94}. In this model
based on methylene "united" atoms, the carbon-carbon bonds and skeletal bending
angles are both rigid with bond lengths set to $d_{cc}=1.54$ angstrom and
bending angle set to $\gamma=109.7$ degrees. These geometrical data are
required to evaluate the contour length $L_c$ of the PE chain with $N$ C-C
bonds according to
\begin{equation}
L_c=N d_{cc} \sin(\gamma/2) \label{eq:LC}
\end{equation}
The intra-molecular potential, written for a chain with N+1
methylene groups, is split into a sum of local contributions and
non local interactions,
\begin{equation}
\tilde U({\un\Gamma}_1)=U^{loc} +U^{nonloc}
\label{eq:pot}
\end{equation}
where
\begin{equation}
U^{loc}=\sum_{i=2}^{N-1} V^{tor}(\phi_i)+\sum_{i=0}^{N-4}v_{LJ}
\left(\mid{\un r}_{i+4}-{\un r}_i\mid\right)
\end{equation}
contains single bond torsion contributions $V^{tor}$ of the "n-butane"
type taken from the "CHARMM" force field \cite{SK92,DLR96} and $1-5$
interactions coping with the pentane effect \cite{Flory}. The non-local
potential part is built as a sum of effective pairwise interactions of
the Lennard-Jones type between united atoms at least fifth neighbor
along the polymer contour,
\begin{equation}
U^{nonloc}=\sum_{i=0}^{N-5}\sum_{j=i+5}^N 4\epsilon^\star\left[\left(
{\sigma^\star\over r_{ij}}\right)^{12}-\lambda\left( {\sigma^\star\over
r_{ij}}\right)^{6}\right].
\end{equation}
The attractive part of the pair interactions is modulated by a factor $\lambda$
allowing the solvent quality to be varied by adjusting the relative weight of
repulsive and attractive interactions. Lennard-Jones parameters for the present
PE model are $\epsilon^\star=0.4301KJ/mol$ and $\sigma^\star=3.74\AA$.

For a given temperature T, our PE chain will be at $\theta$ point for a
particular value $\lambda=\lambda^{\star}(T)$ defined as the
${\lambda}$ value which guarantees that
\begin{equation}
\lim_{N\rightarrow \infty}\mid \left<\left({\un r}_N-{\un r}_0
\right)^2\right>\mid_{\lambda=\lambda^\star} \equiv \lim_{N\rightarrow
\infty}\mid \left<R_0^2\right>\mid_{\lambda=\lambda^\star}\sim N
\end{equation}
leading to the Flory characteristic ratio $C_\infty$ defined by
\begin{equation}
\lim_{N\rightarrow \infty}\frac{1}{(N \ell_{cc}^2)}
\mid\left<R_0^2\right>\mid_{\lambda=\lambda^\star} \equiv
C_\infty(T)
\end{equation}
At T=400K, it was found that $\lambda^{\star}=0.505 \pm0.005$ and
$C_\infty=7.9 \pm0.2$ \cite{DLR96}. This value of $\lambda^{\star}$
differs slightly from the value $\lambda^{\star}=0.53$ quoted in
reference \cite{SMB94} which was obtained at the same temperature with
much shorter chains and a slightly different n-butane torsional form. In
the same work \cite{SMB94}, the temperature dependence of
$\lambda^{\star}$ for the PE model at $\theta$ point was estimated to
be
\begin{equation}
{\lambda^{\star}(T)\over T}={0.2\over T} + 0.0008 K^{-1}\label{eq:theta}
\end{equation}
Performing an equilibrium sampling of the PE chain with the potential
$\tilde U({\un\Gamma}_1)$ yields an energy at fixed ${\un R}$ given by
eq.(\ref{eq:energy}) where
\begin{equation}
{\partial \beta\tilde U\over \partial \beta}=
V^{loc}+\sum_{i=0}^{N-5}\sum_{j=i+5}^N 4\epsilon^\star\left[\left(
{\sigma^\star\over r_{ij}}\right)^{12}-{\partial \beta\lambda^{\star}\over
\partial \beta}\left({\sigma^\star\over r_{ij}}\right)^{6}\right]
\label{eq:split}
\end{equation}
in which, according to eq.(\ref{eq:theta}), ${\partial
(\beta\lambda^{\star})/\partial \beta}=0.2$.

The internal energy of the PE chain can be rewritten as a sum of
potential contributions,
\begin{equation}
E(R,T)=E^{loc}(R,T)+E^{rep}(R,T)+ \frac{\partial \beta\lambda^{\star}}{\partial
\beta} E^{att}(R,T) \label{eq:splitav}
\end{equation}
where the three terms are the local potential contribution, due to torsion plus
the 1-5 interactions, the repulsive and the attractive parts of non local
contributions respectively. They are ensemble averages of the corresponding
terms in eq.(\ref{eq:split}).

\section{Computational methodology}

Single chain Monte-Carlo simulations of PE have been performed by the
configurational bias Monte-Carlo (CBMC) method \cite{FS96}, in combination with
reptation moves of a few methylene units. Such a numerical procedure is
relatively simple as long as the chain is free of any constraint on the end to
end vector. We explain below how the method works for a free chain and how it
can be adapted for sampling a fixed-$f$ chain ensemble, i.e. when a pair of
equal and opposite stretching forces are applied at both chain ends. Using end
to end distance histogram analysis, these fixed stress ensemble samplings are
shown to yield the fixed-$R$ ensemble data which are more appropriate to
discuss the thermo-elastic properties of chains.

\subsection{Sampling in the fixed-f single chain ensemble}

For free chains (not subjected to external forces), a sequence of
$k$ bonds is grown at one end of the chain (the particular end
being chosen at random) while a sequence of $k$ bonds is suppressed
at the other end. Individual dihedral angles (the only relevant
variables) associated with the growth process are generated
according to a distribution $\propto \exp\left(-\beta
V_{tor}\right)$, the rest of the interactions influencing the
acceptance rate through the Rosenbluth weight \cite{FS96}.

When simulating chains with an external stretching force along the z axis
$(0,0,\pm f)$, the associated external potential energy can be taken into
account in the dihedral angle generation by sampling a distribution $\propto
\exp\left(-\beta (V_{tor}-zf) \right)$ where z is the projection of the new
bond in the direction of the external force. Alternatively this external energy
contribution can be taken into account in the Rosenbluth weight. We found that
the latter method works well as long as the force is not too strong
$(f<k_BT/b_k)$ but that it is advantageous to switch to the former method when
the force becomes higher.

For a PE chain of N bonds sampled by MC with acceptance probability $P_{acc}$
for the individual CBMC/reptation moves of $k$ bonds, the generation of a new
independent configuration typically requires $[(N/2k)^2 P_{acc}^{-1}]$ MC
attempted steps \cite{PAR97}. We did experiments for T=400K  using
$\lambda=\lambda^\star(400)=0.505$ for N=256 and 512 which model $\theta$ point
conditions \cite{DLR96}. A few runs in good solvent at the same temperature
were performed with $\lambda=0.01$ for N=256. All results have been produced
with $k=8$ methylene units for the CBMC/reptation moves. A minimum of about
20.000 independent configurations have been generated in each experiment. Such
a large sampling was motivated by the need of a high statistical accuracy in
the histogram analysis to be described below.

\subsection{Fixed-R ensemble results}

Fixed-$R$ ensemble results can easily be obtained from a fixed-$f$
ensemble sampling using histograms based on the instantaneous $R$
value of the sampled stressed chain.

Let us first remind that the fixed-$f$ fixed-$T$ ensemble is
characterized by a partition function $Z_f$ defined as
\begin{equation}
Z_f(f,T)= \int d{\bf r}^N \exp\left[-\beta \left(U({\bf r}^N)-({\bf
r}_N-{\bf r}_0) \cdot {\bf f }\right)\right] \label{eq:Zf}
\end{equation}
with the property that $Z_f(0,T)$ is equivalent to the partition
function of a free chain $Z(T)$ already discussed in eq.
(\ref{eq:histo}). Let $<\hat{O}>_f$ denote the average of an
arbitrary microscopic quantity $\hat{O}$ in that ensemble.

Let us now introduce the probability $W^{f}({\un R},T) d{\un R}$ to
find the end-to-end vector of the chain in the range $({\un R},{\un R}
+ d{\un R})$ in the stressed chain ensemble defined by
eq.(\ref{eq:Zf}). We have
\begin{equation}
W^{f}({\bf R},N)= \frac{\int d{\bf r}^N\delta\left({\bf r}_N-{\bf
r}_0- {\bf R}\right) \exp\left[-\beta \left(U({\bf r}^N)-({\bf
r}_N-{\bf r}_0) \cdot {\bf f }\right)\right]} {Z_f}
\label{eq:distf}
\end{equation}
The distribution of the end-to-end vector in the absence of force,
i.e.~$W^{0}({\bf R},N)$ introduced in eq.~(\ref{eq:histo}), is
related to $W^{f}({\bf R},N)$ by
\begin{equation}
W^{0}({\bf R},N)={Z_f \over Z} \exp\left(-\beta \bf R \cdot \bf f
\right) W^{f}({\bf R},N) \label{eq:bias},
\end{equation}
where the r.h.s. is globally $f$ independent and isotropic in
${\bf R}$ while $W^{f}({\bf R},N)$ depends on $f$ and has a
cylindrical symmetry along the force direction.
By superposing the distributions
$W^{0}$ (known up to some multiplicative constant using eq.
(\ref{eq:bias})) obtained from the histograms of $W^{f}({\bf
R},N)$ established for a set of simulations at several force
intensities, the $W^{0}(R,T)$ profile can be obtained over a large
range of $R$ values with relatively modest computational efforts.

In order to estimate $O({\un R},T)$ as defined in eq.
(\ref{eq:averag}) from the sampling of the stressed chain ensemble,
one exploits the relationship
\begin{equation}
<\hat{O} \delta\left({\bf r}_N-{\bf r}_0- {\bf R}\right)>_{f}=
W^{f}(R,T) <\hat{O}>_R \label{eq:avhist}
\end{equation}
This expression means that $O(R,T) \equiv <\hat{O}>_R$ can be
estimated within the fixed-$f$ ensemble sampling by a simple
arithmetic average of the individual $\hat{O}$ values relative to
all configurations with an end-to-end vector lying in the relevant
box in the $R$ histogram.

In table \ref{table:tab1}, we list the characteristics of the different
fixed-$f$ experiments which are exploited in this paper. In the last column,
for each force value and for the indicated sampling quality, we mention the
range of distances which has been exploited to reconstruct the unperturbed
chain $W^{0}(R,T)$ profile according to eq.(\ref{eq:bias}).

\section{Results}

\subsection{Elastic properties of the realistic PE model}

Figure \ref{fig:ext} shows the force extension law for a $\theta$ chain of PE
at 400K obtained via fixed-$f$ ensemble calculations. The data used are, after
transformation to reduced units, the points $(<{\bf R}>_f,f)$ obtained from
individual simulation runs conducted for a specific combination of $f$ and $N$
values (see table \ref{table:tab1}). To emphasize the universal character (N
independent) of the law, we display the data corresponding to two chain lengths
N=256 and 512 in reduced variables, namely we plot the reduced force $f b_k/k_B
T$ using the equilibrium estimate of $b_k=14.9$ angstrom \cite{DLR96} versus
the reduced elongation $X=R/L_c$ with $L_c$ evaluated according to
eq.(\ref{eq:LC}). The force extension law in the fixed-$R$ ensemble (not shown)
was also estimated using eq.(\ref{eq:fder}) performing the derivative
of a best fit function joining the free energy data of $ln \left(\beta
A(R,T)\right)$(to be discussed later) and the elasticity curve was found to be
in close agreement with the fixed-f ensemble one, except in the $R<R_0$ region
\cite{N86,TPR99} where the two ensembles strongly differ. (Note that the size
of this $R/L_c$ pathological region decreases as $N^{-1/2}$ with chain length).

The data obtained for $N=256$ and $N=512$ yield a unique elasticity curve
$w(X)$ which will be treated in the following as a 'quasi experimental' curve
corresponding to the adopted 'realistic' $\theta$ chain PE model. We observe a
linear regime up to $\approx 35\%$ stretching which is followed by a rapid
increase of the force due to chain FE. The discussion on the validity of simple
chain elasticity models, like those relative to FJC and WLC models, to
reproduce the MC data is postponed to a later section.

We first analyze how the internal energy of the chain and its various
contributions evolve under stretching. According to eq.(\ref{eq:splitav}), the
attractive part of the non local interactions must be weighted with the factor
${\partial (\beta\lambda^\star)/\partial \beta}$ which is quoted to be $0.2$
by Sariban et al. \cite{SMB94} but that we took equal to $0.215$ after the
analysis detailed in section \ref{sec:GS}. To stress the universality of the
curves, we reduce all energetic contributions by a factor $k_BT$ for
dimensional reasons and by a factor $L_c/b_k$ to work with intensive
quantities.

In figure \ref{fig:ener} , we plot the deviations of various energetic
quantities from their unperturbed value ($R=R_0$) in the same thermodynamic
conditions  versus $X=R/L_c$. Data are shown for two chain lengths $N=256$ and
$N=512$. Throughout this paper, the change of a quantity $Y$ with stretching
will be denoted as $\Delta Y(R)$ and defined as $\Delta Y(R) \equiv
Y(R)-Y(R_0)$.

For N=512, the various contributions to the internal energy are shown on the
same figure and we note that those arising from the single bond rotational
energy largely dominate all other terms. The local $1-5$ interactions appear to
have a marginal influence and, more interestingly, the non-local interactions
contribute in a negligible way, in agreement with the $\theta$ point conditions
imposed by the choice $\lambda=\lambda^{\star}$ in our simulations. We come
back on these solvent quality aspects in section \ref{sec:GS} where we show
that good solvent conditions ($\lambda <<\lambda^{\star}$) yield instead, as
expected, a $R$-dependent additional contribution to the internal energy
$E(R,T)$ of the chain.

\subsection{Rationalization of PE thermo-elasticity data}

\begin{figure}[H]
\centerline{\psfig{file=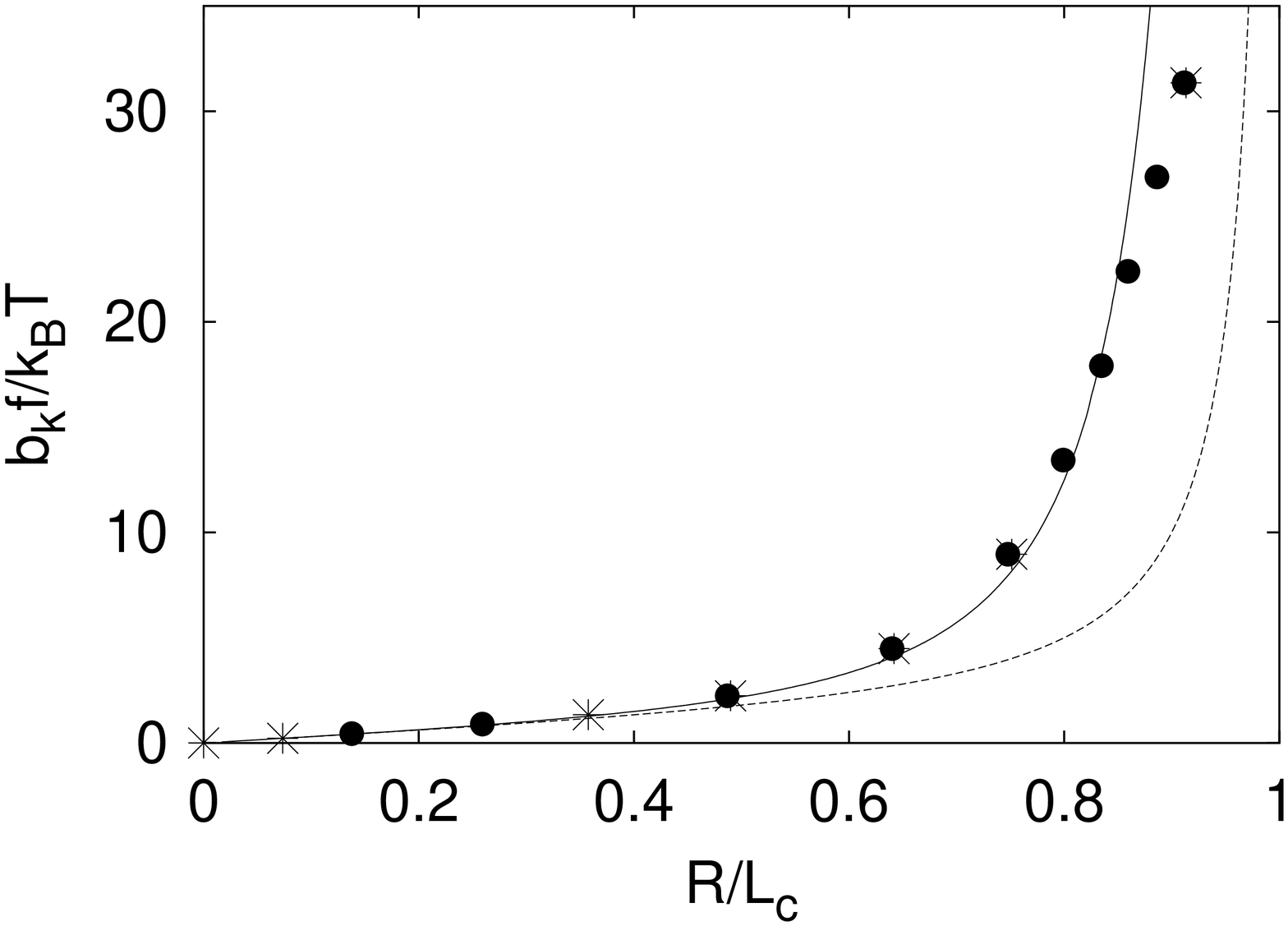,width=8.5cm,angle=-90}}
\caption{Force-extension relationships obtained in the fixed-$f$ ensemble for
N=256 (closed circles) and N=512 (stars). Error bars are smaller than the size
of the symbols. Continuous and dashed lines represent the elasticity law of
the WLC and the FJC models respectively.} \label{fig:ext}
\end{figure}

The elasticity law shown in fig.\ref{fig:ext} for our realistic model of
$\theta$ chains of PE at 400K is a universal curve we will denote as $w(X,T)$.
In the same figure, we compare these data with the functional forms relative to
the FJC and of the WLC models which we denote respectively as $w_{FJC}(X)$ and
$w_{WLC}(X)$. $w_{FJC}(X)$ is simply the inverse Langevin function while the
WLC curve $w_{WLC}(X)$ and its integral $ W_{WLC}(X)$, which are required for
the free energy calculation, were evaluated by the numerical procedure,
accurate to $1.5\%$, which is reported in eqs.(13)-(14) of ref.\cite{MS95}. We
want to emphasize that the curves in figure \ref{fig:ext} are obtained using
the values of $L_c$ and $b_K$ (or $\ell_p$) obtained in the MC model at
equilibrium and are not fitted to MC data. We note that the function $w(X,T)$
based on MC data and both simple models curves are all starting as $\approx 3X$
at small $X$ and diverge (in different ways) as $X$ gets close to $1$.

The origin of elasticity in FJC and WLC models, and by extension to arbitrary
models satisfying eq.(\ref{eq:models}), is often considered as purely entropic,
a character which is commonly interpreted as the result of the reduction of the
number of conformations as the chain extends. This is coherent with the
functional form of the eq. (\ref{eq:models}) as long as the Kuhn segment $b_k$
(or equivalently the persistence length $\ell_p$) is constant, i.e. independent
of temperature and external force. Eq.(\ref{eq:models}) then expresses that $f$
is proportional to $T$ at fixed relative extension, which directly implies that
the force is purely entropic (see eq.(\ref{eq:fe})). The absence of any
energetic contribution to the force is thermodynamically equivalent to the
independence of the internal energy respect to chain stretching, as indicated
by eq.(\ref{eq:fe}). When analyzing together fig. \ref{fig:ener} and fig.
\ref{fig:ext}, it appears that the variation of internal energy with stretching
for PE cannot be avoided as soon as the finite extensibility regime is entered.

In the following, we show that PE elasticity can be modeled over a wide range
of stretching including a large part of the finite extensibility regime on the
basis of eq. (\ref{eq:models}) where $b_k$ is treated as a temperature
dependent quantity. Such a temperature dependence of the persistence length (or
Kuhn segment) has been invoked earlier in the case of the WLC model for DNA
when it models a uniform elastic rod characterized by two parameters, namely
the contour length $L_c$ and the elastic bending modulus $\kappa'$
\cite{ABC97}. In this model, the persistence length, given by
$\ell_p=\kappa'/k_BT$, decreases with increasing temperature as a result of
increasing Brownian undulations of the flexible rod. Here, for synthetic
polymers, the temperature dependence of $b_k$ or $l_p$ must be reflected on the
unperturbed size of the free polymer $R_0$ at $\theta$ point, given the
relationships $R_0^2=L_c b_k=2 L_c \ell_p$.

The form of eq. (\ref{eq:models})with a temperature dependent persistence
length has some general implications on the elasticity law. If we denote by
$W(X)$ the integral between $0$ and $X$ of $w(X)$, we easily get the free
energy and internal energy deviations from the unstressed state
\begin{eqnarray}
\Delta A(R)&=&\frac{L_c k_BT}{\ell_p} W(R/L_c)\\ \Delta E(R)&=&\frac{L_c
k_BT}{\ell_p} \frac{d ln(\ell_p)}{d ln(T)} W(R/L_c) = \frac{d ln(\ell_p)}{d
ln(T)} \Delta A(R) \label{eq:mod}
\end{eqnarray}
Eq. (\ref{eq:mod}) indicate that $\Delta E(R)$ and $\Delta A(R)$ are
proportional. Using eqs. (\ref{eq:fder}) and (\ref{eq:fe}), the proportionality
factor satisfies
\begin{equation}
\frac{d ln(\ell_p)}{d ln(T)}= {\Delta E(R)\over \Delta A(R)}= \frac{f_E}{f}
\label{eq:ratio}
\end{equation}
which implies that the enthalpic/entropic partition of the total force is
independent of stretching.

In figure \ref{fig:enlib}, eq.(\ref{eq:mod}) is tested for two chain lengths
($N=256$ and $N=512$). We indeed find a proportionality between the free energy
change and the corresponding internal energy change over a stretching range
going up to about $65\%$, the best value for the proportionality factor being
$d ln(\ell_p)/d ln(T)=-0.42$. Experimental values of $d ln(R_0^2)/dT$ for PE in
different $\theta$ solvents in the temperature range (373K-463K) are negative
with modulus in the range $(1.0\div 1.2) 10^{-3}K^{-1}$\cite{Flory}. At $400K$,
this correspond to values of $d ln(R_0^2)/d ln(T)=d ln(\ell_p)/d ln(T)$ in the
range $-0.48\div-0.40$ in agreement with our result. Previous calculations for
PE at equilibrium performed at various temperatures on a very similar model
lead to $d ln(R_0^2)/d ln(T)=-0.5\pm 0.1$\cite{SMB94} in agreement with our
finding based on stretching experiments.

The results shown in figure \ref{fig:enlib} lead to several important
conclusions. The elasticity law of the PE model at $\theta$ state is well
represented by a law of the type as in eq. (\ref{eq:models}), with a
temperature independent r.h.s., up to a relative extension of $65\%$. The
variation of the internal energy in this range of elongations is provided by
eq. (\ref{eq:mod}). This elasticity law implies that, in the range of
elongations below $65\%$ which includes a large fraction of the finite
extensibility regime, one has indeed constant ratios $f_E/f =-0.42$ and
$f_S/f=1.42$. Beyond this regime of extension, the more general form of
the elasticity law given in eq. (\ref{eq:finex}) with a temperature dependent
Kuhn segment, must be considered. Note that simple ideal model like FJC and
WLC, even with temperature dependent Kuhn segment or persistence length, cannot
provide elasticity laws of this kind.

The above picture on the thermo-elasticity of simple chain models such as PE
reveals that, as the temperature decreases, the average size of the chain
increases as a result of the relative increase of the population of trans
conformers. In fig.\ref{fig:ng}, we show the variation with stretching of the
gauche population (expressed in percent) $\Delta n_g(R)=n_g(R)-n_g(R_0)$ with
respect to the gauche population in the unstretched value $n_g(R_0)=65\%$, at
T=400K. This curve has been obtained by the histogram outlined in eq.
(\ref{eq:avhist}) on the basis of the series of fixed-$f$ simulations for
$N=512$ listed in table \ref{table:tab1}. The stretching variable used for the
abscissa is chosen to test the Abe-Flory prediction $\Delta n_g(R)= k
((R/R_0)^2-1))$ with k=-0.656 which is represented as a straight line in
fig.\ref{fig:ng} \cite{AF70}. We observe that Flory's theory predicts quite
well the population shift up to $R \approx 3 R_0$ (corresponding to $X=R/L_c
\approx 0.45$ for this chain length). The energetic part of the force clearly
comes from the increased tendency for local bonds to go to lower torsional
energy (trans state) as R increases.

\subsection{The superiority of the WLC model}

Data displayed in fig.\ref{fig:ext} and fig. \ref{fig:enlib} show respectively
the variation of the reduced force $w(X)$ and the reduced free energy $W(X)$ as
a function of $ X=R/L_c$ for the two PE chain lengths investigated. In
fig.\ref{fig:ext}, we show the predicted curves $w_{FJC}(X)$ and $w_{WLC}(X)$
corresponding respectively to the two basic chain models while in fig.
\ref{fig:enlib}, only $W_{WLC}(X)$ is indicated. It turns out that, for the
present PE case, the intermediate range of stretching is much better
represented by the WLC model as it matches 'experimental' data up to a relative
stretching of $\approx 85\%$ for the force-extension law. At larger stretching,
we note that the limiting behavior of the WLC model elasticity law $(L_c-R)
\propto f^{-1/2}$ gives a too slow asymptotic evolution toward $R=L_c$. The
same qualitative picture of elasticity has been observed in experiments on
double-stranded DNA\cite{MS95,ABC97}. We have however already noticed as the
internal energy variation with stretching limits the validity of EOS of the
type predicted by ideal models to the regime  $X \in [0,0.65]$, narrower than
is inferred by the force-extension behavior alone. This consideration puts
some doubts on the present interpretation of the experimental elasticity laws
through ideal models like FJC and WLC. We suggest that the validity of such
models must be supported by internal energy versus extension curves.

In the interpretation of experimental data for DNA, the regime of extreme
stretching ($X\ge 0.9$) has been explained invoking a stretching dependent
contour length $L_c(f)=L_c^0(1+f/f_0)$, where $L_c^0$ is the contour length of
the unstretched chain and $f_0$ is the elastic modulus of the
chain\cite{Odijk95,RFG98,JNO00,Merkel01}, to be used in connection with the
elasticity law of the WLC model. The addition of the new parameter $f_0$, and
its possible temperature dependence, allows the extended WLC model to overcome
its previous limitation of predicting a constant $f_e/f$ ratio\cite{nota}.
Although the qualitative feature of the departure of our MC data from the
primitive WLC model prediction is very similar to that observed for
double-stranded DNA, the elastic WLC model is of little relevance here since
the contour length of our PE chain model, with rigid bonds and fixed bending
angles, is constant.

We could instead define a stretching dependent persistence length $\ell_p(T,X)
\equiv \ell_p^0(T)(1+\Gamma(X,T))$, where $\ell_p^0(T)$ is the persistence
length of the unstretched chain and $\Gamma(X,T)$ a generic function, to be
used to represent the MC data in terms of the ideal WLC model
\beq
\frac{2f\ell_p^0(T)}{k_BT}=w(X,T)= \frac{w_{WLC}(X)}{(1+\Gamma(X,T))} 
\eeq
Note that the temperature dependence of $\Gamma$ is necessary in order to
obtain the observed deviation from a constant value for the ratios $\Delta
E(X)/\Delta A(X)$ and $f_E/f$. Attempting this analysis with our
data, we find that $\Gamma(X,T)$ remains very close to zero up to $X\simeq
0.6$, is slightly negative between $X=0.6$ and $X=0.85$ and then diverges to
$\infty$ for larger $X$. Although in the intermediate regime, the temperature
derivative of $\Gamma(X,T)$ could be invoked to explain the thermoelasticity
behavior of our PE chain, the divergence of $\Gamma(X,T)$ at large X
demonstrates the inadequacy of the WLC at very high stretching.

\section{A few remarks on the solvent effects on the
thermo-elasticity properties}\label{sec:GS}

In order to investigate the solvent quality effect on PE, we performed a
series of experiments for N=256 at 400K, but using $\lambda=0.01 <<
\lambda^{\star}(400)$ with the aim to model good solvent conditions at the same
temperature. Excluded volume forces here modify the elasticity law as we showed
earlier \cite{TPR99}.

\begin{figure}[H]
\centerline{\psfig{file=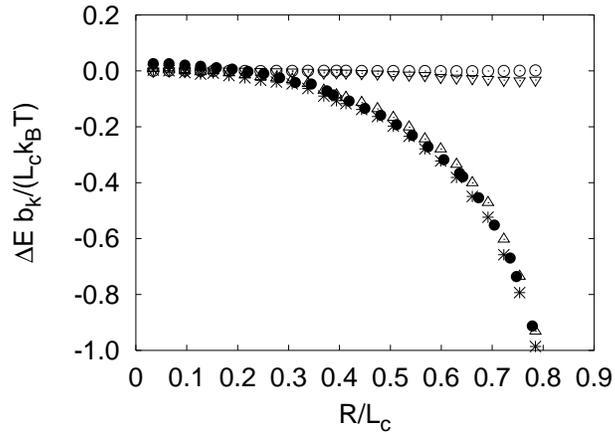,width=8.5cm,angle=-90}} 
\caption{Reduced internal energy difference $\Delta E b_k/L_c k_B T$ 
versus the relative extension $X=R/L_c$ for a
PE chain at $\theta$ point (400K) of N=256 (closed circles) and N=512 (stars).
For N=512, we also show the various contributions to the reduced internal
energy namely the torsional energy (upward triangles) the 1-5 local
interactions (downward triangle) and the non local interaction (open circles).}
\label{fig:ener}
\end{figure}

\begin{figure}[H]
\centerline{\psfig{file=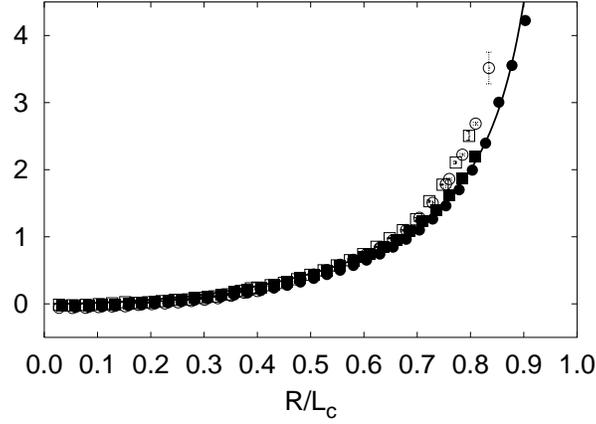,width=8.5cm,angle=-90}} 
\caption{Comparison
between $\Delta A(R)(\ell_p/L_c k_B T)$ (closed symbols) and 
$\Delta E(R)(\ell_p/L_c k_B T)(d ln(\ell_p)/d ln(T))^{-1}$ (open symbols) for
N=256 (circles) and N=512 (squares) versus the relative stretching $R/L_c$. The
value $[d~ln(\ell_p)/d~ln(T)]=-0.42$ has been adopted to rescale the
internal energy difference. The continuous curve represents $W_{WLC}(x)$ as
defined in the text.} \label{fig:enlib}
\end{figure}

\begin{figure}[H]
\centerline{\psfig{file=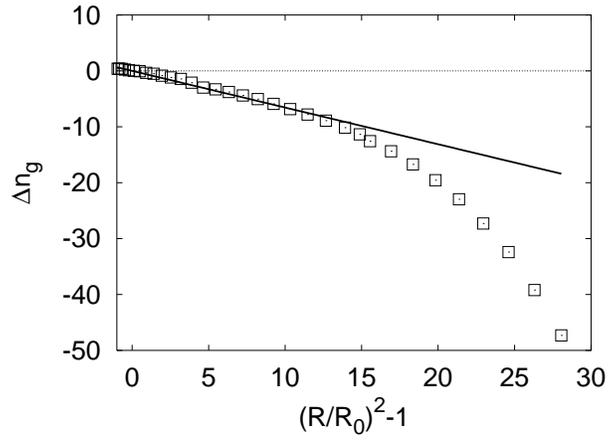,width=8.5cm,angle=-90}} 
\caption{Percentage of
gauche population with respect to the unstretched state $\Delta
n_g=n_g(x)-n_g(0)$ in the fixed-$R$ ensemble as a function of $(R/R_0)^2-1$ for
a N=512 chain at $\theta$ point. The straight line is the Flory's prediction
for small stretching.} \label{fig:ng}
\end{figure}

\begin{figure}[H]
\centerline{\psfig{file=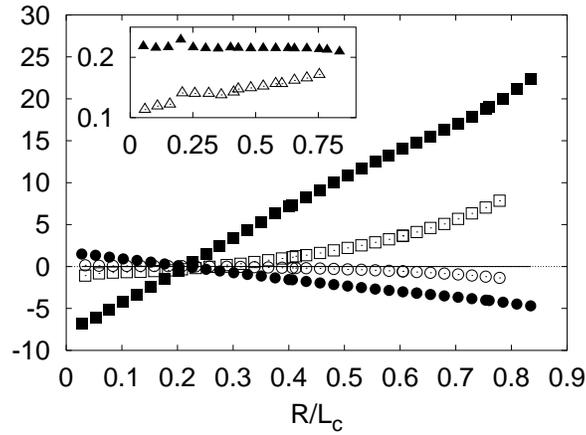,width=8.5cm,angle=-90}} 
\caption{Attractive
(squares) and repulsive (circles) contributions of the non local part of the
energy with respect to the unstretched state in good solvent (open symbols) and
$\theta$ solvent (closed symbols) at 400K. Energies are normalized to $k_BT$.
In the inset we show the ratio of minus the repulsive part to the attractive
part which is constant for $\theta$ solvent but depends on stretching for good
solvent.} \label{fig:comp}
\end{figure}

In figure \ref{fig:comp}, we compare the $R$ dependence of the non local part
of the internal energy for good and $\theta$ solvents. We already noticed in figure 
\ref{fig:ener} that the sum of the repulsive and attractive energy contributions arising 
from non-local interactions is $R$
independent at $\theta$ point, a behavior we now show to be the result of
subtle cancellation effects. Figure \ref{fig:comp} shows the deviations $\Delta
E^{rep}(R,T)$ and $\Delta E^{att}(R,T)$ as defined in eq.(\ref{eq:splitav}) for
the different solvent conditions. While at $\theta$ point, both terms exhibit
linear dependence on $R$, the dependence is non linear in good solvent. These
behaviors imply that the ratio $[-\Delta E^{rep}(R,T) /\Delta E^{att}(R,T)]$, 
shown in the inset of figure \ref{fig:comp}, is constant in the
whole range of stretching in $\theta$ solvent only. Therefore, this constant
value $0.215$ can be taken as an estimate of 
${\partial(\beta\lambda^{\star})/\partial \beta}$ 
which guarantees that the non local
part of the internal energy does not depend on stretching. We note that this
value used to produce data in figure \ref{fig:ener} is in agreement with the
prediction of Sariban et al. \cite{SMB94}.

In good solvent, the contributions of repulsive and attractive non-local forces
do not compensate anymore, whatever value is adopted for 
$\partial(\beta\lambda)/\partial \beta$. Assuming that $\lambda$ is nearly
temperature independent in good solvent conditions,  
$\partial (\beta\lambda)/\partial \beta\approx \lambda$ which provides a monotonous
decrease of internal energy with stretching (not shown). This indicates that
dominant repulsive non local interactions disappear progressively as the size
of tensile blobs characterized by self-avoiding walk statistics reduces with
increasing force \cite{dG79}.

\section{Conclusions}

On the basis of a realistic chain model of polyethylene with solvent mediated
non local interactions, the thermo-elastic properties of an individual chain at
$\theta$ point and in good solvent conditions have been investigated by
Monte-Carlo simulations sampling both fixed-$f$ and fixed-$R$ ensembles over a
large stretching regime.

The $\theta$ state was previously obtained by adjusting an
effective parameter in the attractive part of the non-local interactions so
that the chain size grows as $N^{1/2}$ at large $N$\cite{SBR92,SMB94,DLR96}. We
have shown in section 6 that for such $\theta$ point conditions, the non-local
interactions do not contribute to the average internal force over the whole
stretching range. This property could be taken as an alternative criterion
(using a single chain length) to search for $\theta$ point conditions for a
particular polymer+solvent model. In good solvent conditions, non local
interactions do contribute to the internal force for all stretching values.

We found that the thermo-elastic properties of the chain at $\theta$ point
around 400K in terms of $X=R/L_c$ are quite well represented, up to $X=65\%$,
by an EOS where the temperature and relative elongation dependencies are
factorized, namely $f(X,T)= k_BT w(X)/\ell_p(T)$, where $\ell_p$ is the 
persistence length of the chain determined 
from free chain simulations\cite{DLR96}. 
From the variation of the internal energy and free energy differences with stretching,
we inferred the value of the temperature derivative of $\ell_p$, namely 
$d ln(\ell_p) / dln(T)=-0.42$ in fair
agreement with experimental observations and theoretical predictions for PE
chains at $\theta$ point. We point out here that this temperature variation
implies that the internal force at any stretching has a non negligible
enthalpic part which is negative because extension drives more and more
dihedral angles to adopt the lowest energy trans state.

Up to $65\%$, the $X$ dependencies of the force, i.e. the function $w(X)$ above
(see eq. (\ref{eq:models})), and the of internal energy difference (see eq. (\ref{eq:mod})) 
turn out to be very well represented by the WLC model, using the contour length and
persistence length of our atomistic PE model at equilibrium, without recourse
to any adjustable parameter. By comparison, the FJC model appears to be much
less representative of the PE finite extensibility effects. This result may be
useful to interpret AFM stretching experiments for which the minimal force
which can be detected is above the crossover force between the scaling regime
and the finite extensibility regime. We stress again here that the validity of
the WLC model, in the present context, does not imply that the internal force
has a purely entropic character.

Ad-hoc extensions of ideal chain models which have been proposed so far in the
literature do not seem to apply to the highest stretching regime explored here
($65\%<X<90\%$). The general question of thermodynamic consistency of these
"elastic" WLC and FJC models should be carefully analyzed before they could be
considered something more than simple fitting functions. In particular, if
stretching is measured at a single temperature, the elasticity law $f(X,T)$ and
the internal energy $E(X,T)$ must be simultaneously considered because thermodynamic
consistency (see eq.(\ref{eq:fe}))directly relates the
change of the internal force with temperature (at constant $X$) to the
change of internal energy with stretching (at constant $T$).

\section{Acknowledgment}
J.T.T. gratefully acknowledges financial support from the BOF-NOI UA 2001 and the Flemish Region-I.W.T.

\newpage

\section {Figures and tables captions}

\newpage

\newpage

\newpage

\newpage

\newpage

\begin{table}[H]
\begin{center}
\caption{List of MC experiments in the fixed-$f$ ensemble for two different
system sizes $N$ at $\theta$ point. The last column gives the range of
extensions statistically accessible, for the given imposed force $f$ and number
of statistically independent configurations of the chain.}
\hskip 1cm

\begin{tabular}{|c|c|c|c|}
\hline
$N$&imposed $f$& \# indep. configurations& range of R/Lc exploited\\
\hline \hline
$256$&$0.0$&$ 73100$&$0.00-0.25$\\
\hline
     &$0.2$&$180400$&$0.25-0.47$\\
\hline
     &$0.5$&$153500$&$0.47-0.62$\\
\hline
     &$1.0$&$95400$&$0.62-0.75$\\
\hline
     &$2.0$&$27000$&$0.75-0.81$\\
\hline
     &$3.0$&$27400$&$0.81-0.87$\\
\hline
     &$4.0$&$25000$&$0.87-0.92$\\
\hline
$512$&$0.0$&$36000$&$0.00-0.28$\\
\hline
     &$0.3$&$35000$&$0.28-0.51$\\
\hline
     &$0.5$&$23000$&$0.51-0.60$\\
\hline
     &$1.0$&$24000$&$0.60-0.71$\\
\hline
     &$2.0$&$19200$&$0.71-0.81$\\
\hline
\end{tabular}
\label{table:tab1}
\end{center}
\end{table}

\newpage

\section{Figure captions}
\begin{itemize}
\item[]{\bf Figure 1}: Force-extension relationships obtained in the fixed-$f$ ensemble for
N=256 (closed circles) and N=512 (stars). Error bars are smaller than the size
of the symbols. Continuous and dashed lines represent the elasticity law of
the WLC and the FJC models respectively.

\item[]{\bf Figure 2}: Reduced internal energy difference $\Delta E b_k/L_c k_B T$ 
versus the relative extension $X=R/L_c$ for a
PE chain at $\theta$ point (400K) of N=256 (closed circles) and N=512 (stars).
For N=512, we also show the various contributions to the reduced internal
energy namely the torsional energy (upward triangles) the 1-5 local
interactions (downward triangle) and the non local interaction (open circles).

\item[]{\bf Figure 3}: Comparison
between $\Delta A(R) (\ell_p/L_c k_B T)$ (closed symbols) and \\
$\Delta E(R)(\ell_p/L_c k_B T) (d ln(\ell_p)/d ln(T))^{-1}$ (open symbols) for
N=256 (circles) and N=512 (squares) versus the relative stretching $R/L_c$. The
value $[d~ln(\ell_p)/d~ln(T)]=-0.42$ has been adopted to rescale the
internal energy difference. The continuous curve represents $W_{WLC}(x)$ as
defined in the text.

\item[]{\bf Figure 4}: Percentage of
gauche population with respect to the unstretched state $\Delta
n_g=n_g(x)-n_g(0)$ in the fixed-$R$ ensemble as a function of $(R/R_0)^2-1$ for
a N=512 chain at $\theta$ point. The straight line is the Flory's prediction
for small stretching.

\item[]{\bf Figure 5}: Attractive
(squares) and repulsive (circles) contributions of the non local part of the
energy with respect to the unstretched state in good solvent (open symbols) and
$\theta$ solvent (closed symbols) at 400K. Energies are normalized to $k_BT$.
In the inset we show the ratio of minus the repulsive part to the attractive
part which is constant for $\theta$ solvent but depends on stretching for good
solvent.

\end{itemize}

\begin{thebibliography}{10}
\bibitem{nota}
A force dependent contour length in eq.(\ref{eq:finex}) or
eq.(\ref{eq:models}), although physically soundable, transform the equations in
implicit ones for the force. From these equation it is difficult to infer the
thermodynamics. We could however think to a generic temperature dependence of
the contour length through the temperature dependence of the extra parameter.
For this elastic WLC model with temperature dependent persistence length
$\ell_p(T)$ and contour length $L_c(T)$ we have the following EOS
$$
\frac{2f\ell_p(T)}{k_BT}=w_{WLC}(X(T))
$$
which implies the following relation for the thermoelasticity
$$
\frac{\Delta E}{k_BT}=\left[\frac{d ln\ell_p}{d lnT}-\frac{d ln L_c}{d lnT}\right]
\frac{\Delta A}{k_BT}+\frac{L_c~d ln L_c}{\ell_p d lnT}~X~w_{WLC}(X(T))
$$
more general than the simple eq.(\ref{eq:mod}).
\end{thebibliography}
\end{document}